# Delicate Ferromagnetism in MnBi$_6$Te$_{10}$

Chenhui Yan[1], Yanglin Zhu[2], Leixin Miao[3], Sebastian Fernandez-Mulligan[1], Emanuel Green[1], Ruobing Mei[2], Hengxin Tan[4], Binghai Yan[4], Chao-Xing Liu[2], Nasim Alem[3], Zhiqiang Mao[2] & Shuolong Yang[1]*

[1]Pritzker School of Molecular Engineering, University of Chicago, Chicago, Illinois 60637, USA
[2]Department of Physics, Pennsylvania State University, University Park, State College, Pennsylvania 16802, USA
[3]Department of Materials Science and Engineering, The Pennsylvania State University, University Park, Pennsylvania 16802, USA
[4]Department of Condensed Matter Physics, Weizmann Institute of Science, Rehovot 7610001, Israel

*yangsl@uchicago.edu

**ABSTRACT:** Tailoring magnetic orders in topological insulators is critical to the realization of topological quantum phenomena. An outstanding challenge is to find a material where atomic defects lead to tunable magnetic orders while maintaining a nontrivial topology. Here, by combining magnetization measurements, angle-resolved photoemission spectroscopy, and transmission electron microscopy, we reveal disorder-enabled, tunable magnetic ground states in MnBi$_6$Te$_{10}$. In the ferromagnetic phase, an energy gap of 15 meV is resolved at the Dirac point on the MnBi$_2$Te$_4$ termination. In contrast, antiferromagnetic MnBi$_6$Te$_{10}$ exhibits gapless topological surface states on all terminations. Transmission electron microscopy and magnetization measurements reveal substantial Mn vacancies and Mn migration in ferromagnetic MnBi$_6$Te$_{10}$. We provide a conceptual framework where a cooperative interplay of these defects drives a delicate change of overall magnetic ground state energies, and leads to tunable magnetic topological orders. Our work provides a clear pathway for nanoscale defect-engineering towards the realization of topological quantum phases.

The first intrinsic magnetic topological insulator $MnBi_2Te_4$ (MBT)[1–16] integrates topology with magnetism, and provides a fertile ground for realizing fascinating topological phases. Importantly, MBT exhibits intralayer ferromagnetism and interlayer antiferromagnetism. The compensated magnetic moments in even-layer MBT and the uncompensated ones in odd-layer counterparts give rise to axion insulators[1] and quantum anomalous Hall (QAH) insulators[14], respectively. The key to tuning and optimizing these topological phases is to precisely control the interlayer magnetic interactions.

A promising route is to construct $MnBi_{2n}Te_{3n+1}$ superlattices, where $Bi_2Te_3$ (BT) buffer layers are inserted between the MBT layers. Previous studies have revealed that superlattices with n ≤ 3 exhibit interlayer antiferromagnetism,[17–23] whereas those with n > 3 display ferromagnetism.[23–25] On the other hand, for higher order superlattices the interlayer magnetic interactions become progressively weaker,[23] which puts a strong limit on the tunability of magnetic interactions and subsequently on the onset temperatures of the resulting topological phases. Notably, previous studies on powder $MnBi_6Te_{10}$[26] or hydrostatically pressurized $MnBi_6Te_{10}$[27] revealed the possibility of tunable magnetic ground states, yet the nature of these samples disallowed angle-resolved photoemission spectroscopy (ARPES) to directly determine the band topology.

An alternative route is to explore the prevalent disorder effects in MBT and related compounds,[28–33] and potentially use disorder to control the interlayer magnetic interactions. Recent experiments on $MnSb_2Te_4$ (MST) and Sb-doped $MnBi_{2n}Te_{3n+1}$ have demonstrated that the Mn/Sb and Mn/Bi anti-site defects play an important role in tuning the system between antiferromagnetic (AFM) and ferromagnetic (FM) phases.[29,34] However, no direct evidence has been found in the momentum space for the broken-symmetry gap in these materials. Moreover, the topological nature of MST is still under intense debates due to the reduced spin-orbit coupling (SOC) effect.[12,35,36] It is thus an intriguing question whether such disorder-mediated magnetic interactions can be realized in the MBT-derived compounds without compromising the SOC effect, which potentially enables tunable topological phases by a sensitive control of the disorder.

In this letter, we report a delicate FM topological insulator state in $MnBi_6Te_{10}$, which is attributed to disorder-mediated magnetic interactions. We employ high resolution laser-based ARPES to detail the electronic structures of the FM and AFM phases, respectively. In the FM phase, a broken-symmetry gap is unambiguously observed on the topological surface state (TSS) of the MBT termination, with a gap onset temperature coinciding with the Curie temperature. In stark contrast, all terminations of the AFM phase of $MnBi_6Te_{10}$ exhibit negligible energy gaps on the TSS. Furthermore, our structural and magnetic characterizations reveal that Mn vacancies in MBT layers and Mn migration from MBT to BT layers are prevalent in FM $MnBi_6Te_{10}$. We provide a conceptual framework where a delicate interplay of Mn vacancies and Mn migration leads to the tunable magnetic phases in $MnBi_6Te_{10}$. Our work not only establishes the first unequivocal FM topological insulator phase in $MnBi_6Te_{10}$, but also demonstrates one of the highest FM $T_c$'s (13 K) among all MBT-derived compounds.[24,25,37] The proposed new scheme of disorder-mediated ferromagnetism provides a pathway towards sensitive nanoscale tuning of topological phases and future topological quantum devices.

The $MnBi_6Te_{10}$ single crystals were synthesized through a self-flux method[20]. So far, only the AFM phase was reported in $MnBi_6Te_{10}$, which is the ground state for $MnBi_{2n}Te_{3n+1}$ ($n \leqslant 3$). The FM phase has 0.05 meV higher total energy per Mn atom that may be compensated by introducing certain type of defects.[23,38] To meet this challenge, we increased the growth temperature window by 5 °C for FM samples comparing with the growth temperature for AFM samples in order to create Mn/Bi antisites and Mn vacancies, as both types of defects can weaken the interlayer AFM exchange coupling.

As shown in Figure 1b, the X-ray diffraction (XRD) patterns for the two magnetic phases exhibit the same (00$l$) diffraction peaks in good agreement with previous studies.[20,39,40] Figure 1c displays the zero-field-cooled (ZFC) and field-cooled (FC) magnetic susceptibilities measured with a $c$-axis applied field $H = 100$ Oe. A sharp Λ-like peak at 10.2 K is observed in the magnetic susceptibility of AFM MnBi$_6$Te$_{10}$, which indicates an AFM transition.[20,23,25] The ZFC-FC bifurcation below 8 K seen in the AFM sample can be attributed to the evolution from a long-range AFM order to a cluster spin glass state,[23] which is likely driven by the magnetic frustration caused by competing FM and AFM interactions. In contrast, the magnetic susceptibility for the FM material shows signatures of a typical FM transition: a rapid increase and plateauing of the susceptibility below the Curie temperature of 13 K. For FM MBT-derived materials, magneto-optical imaging has demonstrated that the ZFC-FC bifurcation in magnetic susceptibilities is likely due to the movement of FM domains.[41] The FM phase is further confirmed by the isothermal magnetization curves in Figure 1e: a typical hysteresis loop expected for FM materials is revealed. In contrast, a spin-flop-like transition is observed in the AFM phase (Figure 1d), in agreement with previous reports on AFM MnBi$_4$Te$_7$.[19]

High-resolution ARPES results on FM MnBi$_6$Te$_{10}$ are presented in Figure 2. Three possible terminations are expected after cleaving, denoted by the top layer as MBT, single BT (1-BT), or double BT (2-BT) terminations. It is crucial to employ micron-scale laser beams to distinguish different terminations. Figure 2 presents three types of electronic structures found on FM MnBi$_6$Te$_{10}$. First, we point out that the termination assignment cannot be based on direct comparison between ARPES data and first-principles calculations, as the latter is often unable to properly model the impacts of defects and quantum confinement in $MnBi_{2n}Te_{3n+1}$ materials.[16,40,42] A common spectroscopic feature characteristic of the MBT termination for all $MnBi_{2n}Te_{3n+1}$ superlattices is a Dirac surface state hybridized with parabolic Rashba bands.[16,40,43] Hence, we associate the electronic structure in Figure 2a with the MBT termination. We assign the spectrum in Fig. 2d to the 1-BT termination due to its strong resemblance with the counterpart on the 1-BT termination of FM $Mn(Bi_{0.85}Sb_{0.15})_4Te_7$.[37] Finally, since our material is phase-pure as demonstrated by the transmission electron microscopy measurement (Fig. 3), we can only assign the last type of ARPES spectrum (Fig. 2g) to the 2-BT termination. The two different types of BT-derived terminations exhibit electron (Figure 2d) and hole (Figure 2g) dopings, respectively. We notice that the qualitative trend is consistent with the previous ARPES data on the BT-derived terminations of AFM $MnBi_6Te_{10}$[40] and FM $MnBi_8Te_{13}$[25], where electron doping systematically decreases as the number of BT layers increases. Nevertheless, the doping change from the 1-BT to the 2-BT terminations in FM $MnBi_6Te_{10}$ is much more dramatic, which may be due to its specific defect configuration [Supporting Information (SI) Note 1].

We focus on the MBT and 1-BT terminations as their Dirac points are clearly resolvable below the Fermi level. Circular dichroism in ARPES measurements exhibits antisymmetric patterns near the Dirac point for both terminations (SI Figure 1). A helical circular dichroism (CD) pattern in ARPES has been successfully used to identify the topological surface state in $MnBi_{2n}Te_{3n+1}$ compounds.[11,43] For the MBT termination, an energy gap of ~15 meV at the Dirac point is visible in the ARPES data at 7.5 K and this gap disappears at 20 K (Figures 2a, 2b). The comparison of energy distribution curves (EDCs) taken at $\bar{\Gamma}$ further highlights this temperature-dependent energy gap: a spectral peak at -0.18 eV at 20 K evolves into a dip upon cooling to 7.5 K. A temperature-dependent band dispersion analysis on the MBT termination yields a gap-closing temperature at 12.7 ± 1.4 K (SI Note 2 and SI Figure 2), yet we cannot completely rule out the existence of a finite gap above $T_c$. For the 1-BT termination, the spectra taken at both temperatures exhibit gapless Dirac cones (Figures 2d-2f).

AFM MnBi$_6$Te$_{10}$ exhibits three types of ARPES spectra on three terminations (SI Figure 3 and SI Figure 4).[21,40,44,45] The energy gaps at the Dirac points are negligible on all terminations of AFM MnBi$_6$Te$_{10}$, consistent with previous reports.[22,40,43] While it is not the focus of our work, we point out that a likely scenario for the gapless TSS on the MBT termination of AFM MnBi$_6$Te$_{10}$ is that the Mn-Bi antisite defects may cause the TSS wavefunction to be relocated into the space between two adjacent MBT septuple layers.[46] Both the non-magnetic BT layers and the opposite magnetic moments from the two adjacent MBT layers may reduce the magnetic gap.

We thus establish the spectroscopic evidence of a magnetically induced broken-symmetry gap in FM MnBi$_6$Te$_{10}$ for the first time. This gap originates from strong interactions between the TSS electrons and the magnetic moments in the MBT layer. In contrast, the TSS electrons on the 1-BT termination are localized to the top BT layer and spatially separate from the magnetic MBT layer, leading to a gapless Dirac point. Gap opening in the MnBi$_{2n}$Te$_{3n+1}$ compound family has been highly controversial.[2,3,6,7,11,16,22,23,25,40,44,47–50] Specifically on MnBi$_6$Te$_{10}$, a 60 meV gap was reported in the AFM phase[44] in contrast to the results from all other ARPES studies on MnBi$_6$Te$_{10}$[22,23,25,40] including ours. However, this gap persists well above the magnetic ordering temperature and is likely due to extrinsic reasons such as local impurities. On the contrary, our clear spectroscopic gap in FM MnBi$_6$Te$_{10}$ which onsets at the magnetic ordering temperature clarifies the physics picture of a broken time-reversal symmetry. In addition, our resolved gap in FM MnBi$_6$Te$_{10}$ is fully consistent with previous observations on FM MnBi$_8$Te$_{13}$.[24,25] The crucial new discovery is that the magnetic phase of MnBi$_6$Te$_{10}$ can be controllably tuned between AFM and FM.

We proceed to investigate the microscopic mechanism leading to tunable magnetic phases. Notably, Mn-doped BT can host ferromagnetism,[51–54] yet the distinct XRD and ARPES results from our FM MnBi$_6$Te$_{10}$ compared to those from Mn-doped BT[51–53] suggest that the proportion of the latter phase in our materials is negligible. Furthermore, structural characterizations by annular dark field scanning transmission electron microscopy (ADF-STEM) do not find any observable impurity phases for AFM or FM samples, as shown in Figure 3. The atomic resolution images exhibit an interleaved structure composed of MBT septuple layers and 2 BT quintuple layers, which is consistent with the crystal structures determined by XRD in Figure 1b. The selected-area electron diffraction (SAED) patterns, determined by the atomic stacking sequences and periodicity along the *c*-axis (SI Figure 5), further confirm the single phase of the FM and AFM MnBi$_6$Te$_{10}$ samples.

Mn vacancies are observed in both AFM and FM MnBi$_6$Te$_{10}$, as highlighted in red in Figures 3a-3c. Strikingly, the concentration of Mn vacancies in the FM samples (Figure 3b) is much higher than that in the AFM samples (Figure 3a). The quantitative percentages of Mn vacancies are hard to determine from ADF-STEM since the intensity is formed by atoms in the projection of the entire atomic columns perpendicular to the imaging plane. The energy dispersive X-ray (EDX) analysis confirms the higher concentration of Mn vacancies in the FM samples (SI Table 1).

Another important type of disorder is Mn migration. This effect is manifested as the ferrimagnetic order induced by the *antisite defects* in MnSb$_2$Te$_4$[28,29] and Sb-doped MnBi$_4$Te$_7$[34,55]. Measurements of magnetic moments at low and high magnetic fields can be used to estimate the density of Mn migration.[32] Using our measured magnetic moments at 0.2 T (Figure 1) and 7 T (SI

Figure 6), we obtain that 8.1% and 11.8% of Mn atoms in FM and AFM MnBi$_6$Te$_{10}$, respectively, have migrated from the original Mn sheets to the neighboring layers. Here we use the chemical formula of Mn$_{1-y-6x}$(Bi$_{1-x}$Mn$_x$)$_6$Te$_{10}$, where $6x$ and $y$ indicate the densities of Mn migration and Mn vacancies, respectively. The comparable densities of Mn migration in FM and AFM MnBi$_6$Te$_{10}$ suggest that the true physical picture for the FM order is more complex than a simple migration-induced ferrimagnetism.[29]

We construct a conceptual model to illustrate that it is the delicate interplay of Mn vacancies and Mn migration that leads to the tunable magnetic phases in MnBi$_6$Te$_{10}$. We consider two Mn sheets where the intralayer and interlayer magnetic interactions are FM and AFM, respectively. An intermediate layer of migrated Mn ions interacts with the original Mn sheets antiferromagnetically. We obtain the energy difference between the FM and AFM alignments of the original Mn sheets (SI Note 3).

$$E_{FM} - E_{AFM} = 2\sum_{i,j}^{(1),(2)} J_{IL,ij} - 2\sum_{i,j}^{(1),(3)} J_{D,ij} = 2H_{IL} - 2H_D \qquad (1)$$

Here $J_{IL,ij}$ is the interlayer magnetic coupling between site $i$ and $j$ in the two original Mn sheets; $J_{D,ij}$ is the defect-induced magnetic coupling across the original Mn sheets and the defect layer. Importantly, the scenarios of $H_{IL} > H_D$ and $H_{IL} < H_D$ lead to the AFM and FM alignment of the two original Mn sheets, respectively.

To evaluate the energy balance in the presence of Mn vacancies and Mn migration, we consider the scaling laws of $H_{IL}$ and $H_D$ in terms of the Mn density in the original sheets ($n_o$) and in the migrated space ($n_m$): $H_{IL} \propto n_o^2$ and $H_D \propto n_o n_m$. These scaling laws are rooted in the microscopic nature of magnetic interactions (SI Note 4), and independent of the numerical details of $J_{IL,ij}$ and $J_{D,ij}$. Subsequently,

$$H_D/H_{IL} \propto n_m/n_o \qquad (2)$$

Eqn. (2) reveals the microscopic mechanism for defect-induced ferromagnetism in MnBi$_6$Te$_{10}$. First, with substantial Mn vacancies in the MBT layer, both $H_{IL}$ and $H_D$ decrease (Figure 3e). The energy balance is determined by $n_m/n_o$. Notably, the saturated magnetic moments at low and high magnetic fields ($M_1$ and $M_2$, respectively) allow us to estimate this ratio[32]: $n_m/n_o = (M_2 - M_1)/(M_2 + M_1)$. Our measurements lead to $n_m/n_o = 0.2$ and 0.13 for the FM and AFM samples, respectively (SI Note 4). The values of $n_m/n_o$ for all the AFM MnBi$_6$Te$_{10}$ materials in the literature are smaller than 0.1.[20,25,38] The > 50% increase of $n_m/n_o$ tips the energy balance between $H_D$ and $H_{IL}$, stabilizing a ferromagnetic phase (Figure 3e). We remark that the more accurate $n_m/n_o$ can be >0.2 for FM MnBi$_6$Te$_{10}$ due to the potentially unsaturated $M_2$ at 7 T.[32]

We emphasize that the simple Ising model neglects other types of magnetic interactions and should only serve as a conceptual framework. Nevertheless, it offers powerful scaling laws under which the delicate adjustment of Mn vacancies and Mn migration leads to tunable magnetic phases in MnBi$_6$Te$_{10}$. We emphasize that this disorder tuning leads to a change of the *global* magnetic ground state, which is to be distinguished from the more trivial disorder effect resulting in local changes of magnetism.[36] Our new insight of quantitatively considering the ratio of $n_m/n_o$ is a substantial advance compared to previous studies which qualitatively pointed out the importance of disorder[36,56] and the theoretical modeling whose predictions sensitively depend on numerical details.[55] This insight, importantly, is obtained by the multi-modal measurements combining magnetization, laser-based µARPES, and atomic-resolution TEM on the same batch of samples. Notably, FM MnBi$_6$Te$_{10}$ exhibits a T$_c$ of 13 K, which is higher than the T$_c$ of 10.5 K for FM MnBi$_8$Te$_{13}$[24,25] and the T$_c$ < 9 K for hydrostatically pressured MnBi$_6$Te$_{10}$[27], suggesting that the disorder-induced ferromagnetism may be more robust than that induced by superlattice stacking or by hydrostatic pressure. The delicate ferromagnetism revealed in this work will serve as a general framework to understand the phenomenology in all MnBi$_{2n}$Te$_{3n+1}$ superlattices.

We hope that our work will serve as a milestone to motivate further theoretical and experimental studies, in particular spectroscopy and microscopy studies with atomic resolutions to resolve what defects form under what growth conditions in disordered FM MnBi$_{2n}$Te$_{3n+1}$. The defect engineering via controlling the growth temperature can potentially be adapted in thin film deposition of MnBi$_{2n}$Te$_{3n+1}$ as well. Understanding how to tune the magnetism and topological states without introducing new chemical elements in the same MBT system through delicately controlling the defects, allows us to selectively realize exotic phases such as the axion insulator state and the quantum anomalous Hall insulator state, paving the road towards functional topological quantum devices at realistic cryogenic temperatures.

**Methods**

**Sample growth and characterization**

MnBi$_6$Te$_{10}$ single crystals were synthesized using the self-flux method,[20] in which Mn, Bi, and Te powders were mixed with a stoichiometric molar ratio and sealed in a carbon-coated quartz tube under high vacuum. For the synthesis of ferromagnetic (FM) MnBi$_6$Te$_{10}$, the mixture was heated to 900 °C in a furnace and held for 10 hours for homogeneous melting. The mixture then underwent a series of cooling and annealing stages: from 900 to 595 °C in 5 hours, from 595 to 580 °C at a rate of 0.1 °C/h, annealed at 580 °C for 48 h, and finally quenched in water at 0 °C. For the synthesis of antiferromagnetic (AFM) MnBi$_6$Te$_{10}$, the same melting was adopted. It was followed by cooling from 900 to 590 °C in 5 hours and then from 590 to 575 °C at a rate of 0.1 °C/h, annealed at 575 °C for 48 h and then quenched in water at 0 °C. The as-grown single crystals were found to be plate-like with luster and lateral dimensions of 2×2 mm$^2$.

The crystallization of the grown single crystals was verified by the sharp (00L) XRD peaks using a Malvern Panalytical Empyrean diffractometer (Cu K$_\alpha$ radiation), as shown in Figure 1b. Magnetization of crystals was measured by the MPMS3 SQUID magnetometer (Quantum Design).

The TEM specimens were prepared using ThermoFisher Helios 660 dual beam system. The specimens were thinned down to electron transparency using 30 kV and 5 kV Ga ion beam and cleaned with 2 kV and 1 kV ion beam. The aberration-corrected STEM imaging was performed using ThermoFisher Titan G2 S/TEM equipped with image and probe correctors. The operating voltage for the STEM imaging was 200 kV. The STEM images have been drift corrected using the non-linear drift correction algorithm. The elemental mapping was acquired with energy dispersive x-ray spectroscopy (EDS) with STEM mode.

**Ultrahigh resolution laser-based angle-resolved photoemission spectroscopy (ARPES)**

Our laser-based ARPES setup was based on a Coherent MIRA Ti:Sapphire oscillator. With a 5 W, 532 nm continuous wave seed laser, the oscillator output > 9 nJ pulses with a central wavelength at 820 nm, a bandwidth of 7.6 nm, a pulse duration of 130 fs, and a repetition rate of 80 MHz. 1 mm beta barium borate (BBO) crystals were used to generate the second harmonic (410 nm) and fourth harmonic (205 nm), the latter of which was used for ARPES measurements. The optical bandwidth of the 6 eV beam was expected to be 2.7 meV due to the finite BBO thicknesses. The overall energy resolution incorporating the ARPES analyzer resolution was characterized as 4 meV.[57] The beam waist at the optical focal point was less than 10 microns.[57]

**The Supporting Information** contains the following supporting contents: Defects and doping, Circular dichroism in the ARPES spectra of ferromagnetic $MnBi_6Te_{10}$, A Side-by-side comparison of FM and AFM $MnBi_6Te_{10}$ using the temperature evolutions of the TSS's on the MBT termination, Electronic structure of antiferromagnetic $MnBi_6Te_{10}$, Circular dichroism in the ARPES spectra of antiferromagnetic $MnBi_6Te_{10}$, Selected area electron diffraction (SAED) patterns of $MnBi_6Te_{10}$, High-field magnetization measurements of AFM and FM $MnBi_6Te_{10}$, Estimates of the chemical concentrations of $MnBi_6Te_{10}$ based on the scanning transmission electron microscopy – energy dispersive x-ray (STEM-EDX) analysis, Magnetic interactions, and the determination of the Mn density ratio between different layers.

**Acknowledgments**

The ARPES work was in part supported by the US Department of Energy (DOE), Office of Science, Basic Energy Sciences, Materials Science and Engineering Division, under contract No. DE-AC02-06CH11357, and in part supported by NSF DMR-2145373. The financial support for sample preparation was provided by the National Science Foundation through the Penn State 2D

Crystal Consortium-Materials Innovation Platform (2DCC-MIP) under NSF cooperative agreement DMR-1539916 and DMR-2039351. Z.Q.M. acknowledges the support from the US National Science Foundation under grant DM-1917579. C. X. L. and R.B.M. acknowledge the support of the U.S. Department of Energy (Grant No. DESC0019064). B.Y. acknowledges the financial support by the Willner Family Leadership Institute for the Weizmann Institute of Science, the Benoziyo Endowment Fund for the Advancement of Science, Ruth and Herman Albert Scholars Program for New Scientists, the European Research Council (ERC) under the European Union's Horizon 2020 research and innovation programme (Grant No. 815869). N.A. and L.M acknowledge the support by NSF through the Pennsylvania State University Materials Research Science and Engineering Center DMR-2011839 (2020 - 2026). L.M and N.A. acknowledge the Air Force Office of Scientific Research (AFOSR) program FA9550-18-1-0277 as well as GAME MURI, 10059059-PENN for support.

**Author contributions**

C.Y., S.M., E.G., and S.Y. performed the ARPES measurements. Y.Z. and Z.M. grew the samples and performed the magnetization measurements. L.M. and N.A. performed the STEM measurements. All authors discussed the results and commented on the paper. C.Y. and S.Y. wrote the paper with input from all authors.

**Notes**

The authors declare no competing interests.

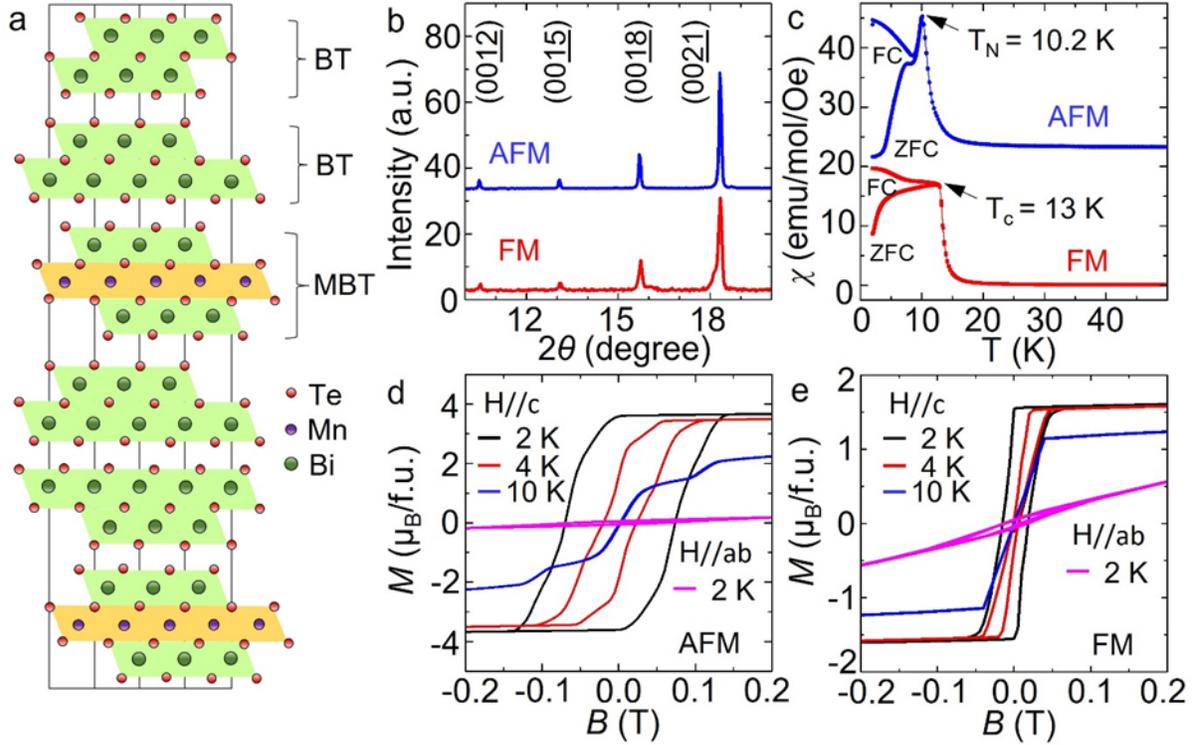

Figure 1. Structural and magnetic characterizations of $MnBi_6Te_{10}$. (a) Schematic crystal structure of $MnBi_6Te_{10}$. (b) X-ray diffraction of ferromagnetic (FM, red) and antiferromagnetic (AFM, blue) $MnBi_6Te_{10}$. (c) Temperature dependent zero-field-cooled (ZFC) and field-cooled (FC) magnetic susceptibilities of FM (red) and AFM (blue) $MnBi_6Te_{10}$ using an external field H = 100 Oe along the *c* axis. The results corresponding to the AFM samples are offset vertically for clarity. (d, e) Isothermal magnetization curves with the magnetic field applied along the *c* axis and in the *ab* plane at various temperatures in (d) AFM and (e) FM $MnBi_6Te_{10}$.

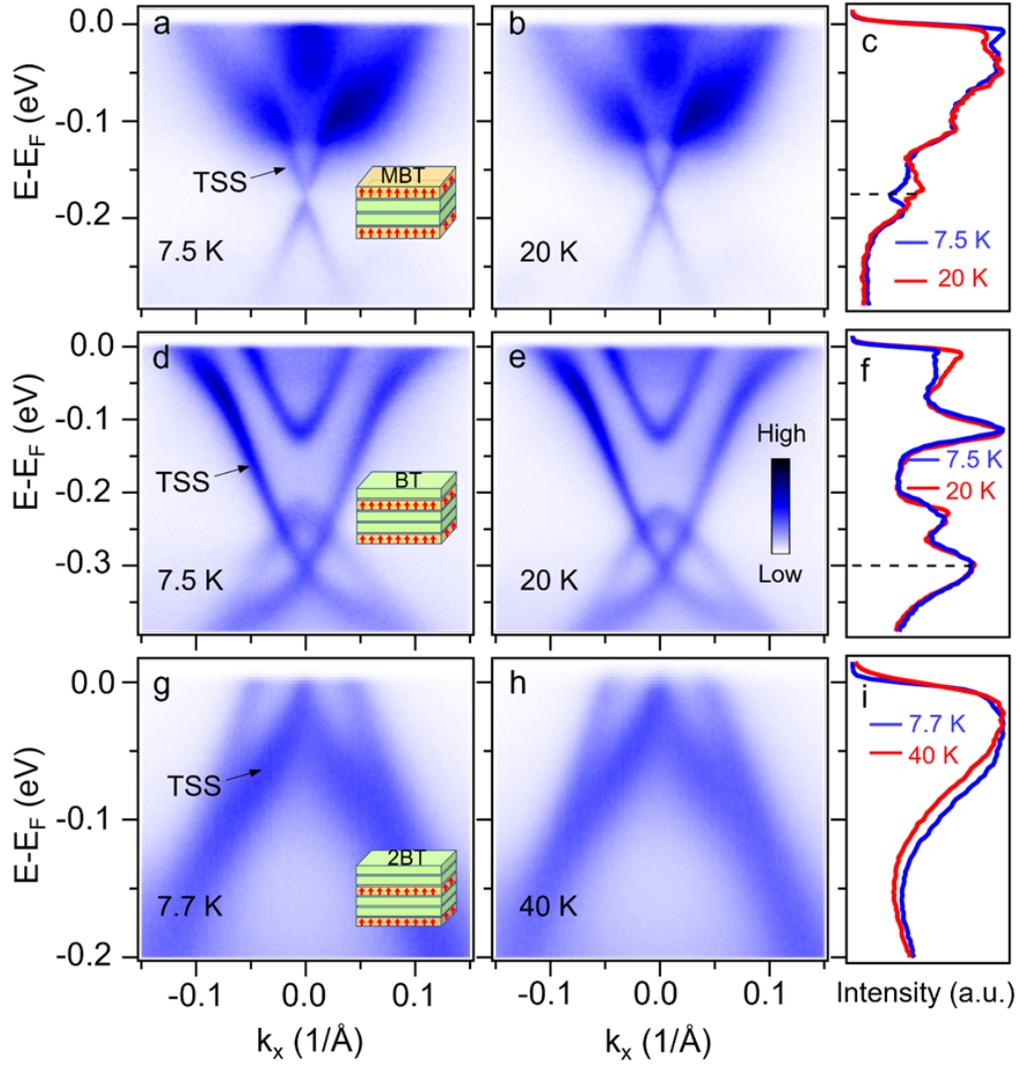

Figure 2. Electronic structure of ferromagnetic MnBi$_6$Te$_{10}$. Energy-momentum spectra along $\bar{\Gamma} - \bar{M}$ at (a) 7.5 K, and (b) 20 K. The insert in (a) illustrates the MnBi$_2$Te$_4$ (MBT) termination. (c) Comparison of energy distribution curves at $\bar{\Gamma}$. An energy gap is observed at the Dirac point (black dashed line) at 7.5 K. The counterpart results for the 1-Bi$_2$Te$_3$ (1-BT) termination are plotted in (d-f): (d, e) energy-momentum spectra, and (f) energy distribution curves at $\bar{\Gamma}$. The counterpart results for the 2-Bi$_2$Te$_3$ (2-BT) termination are plotted in (g-i): (g, h) energy-momentum spectra, and (i) energy distribution curves at $\bar{\Gamma}$.

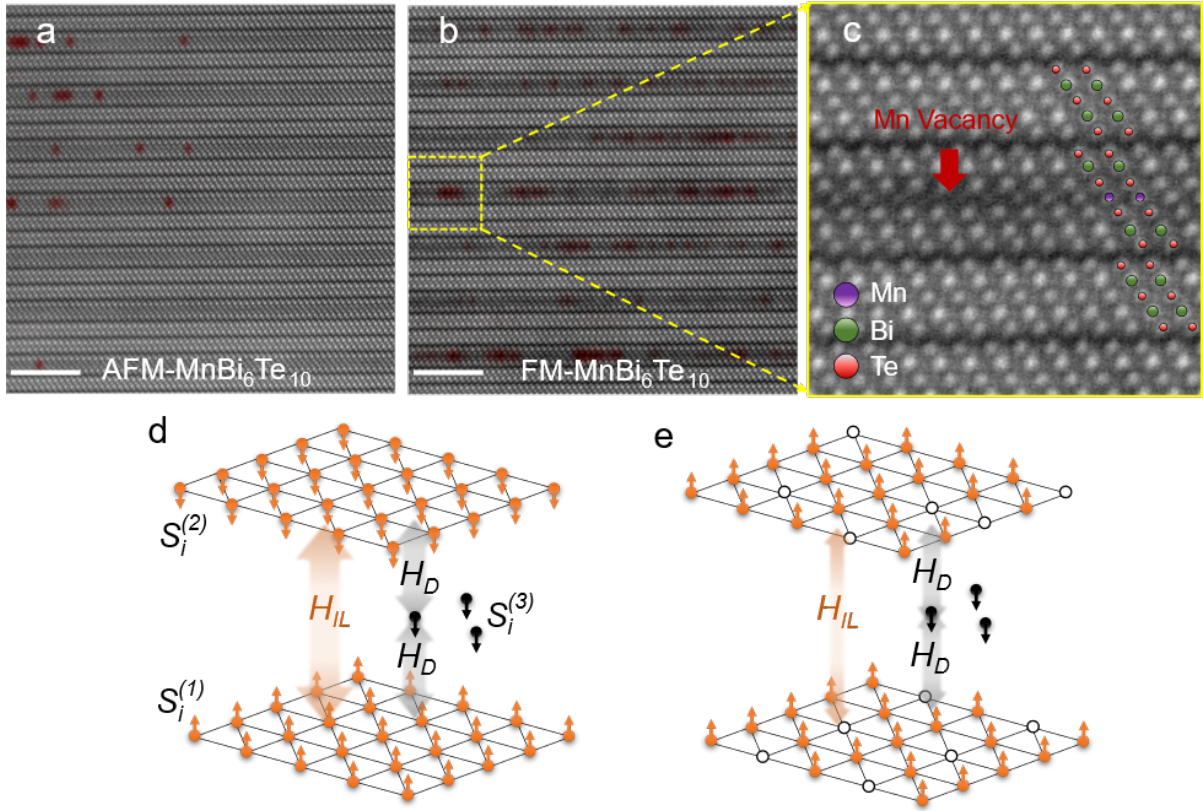

Figure 3. Structural characterizations of MnBi$_6$Te$_{10}$. Annular dark field scanning transmission electron microscopy (ADF-STEM) images of (a) AFM and (b) FM MnBi$_6$Te$_{10}$. Scale bar indicates 5 nm. Mn vacancies are highlighted by red shading, where the intensity for the atomic column of Mn is lower than the value of two standard deviations below the Mn mean intensity. (c) Magnified image (yellow box in (b)). (d) A cartoon illustration showing $H_{IL} > H_D$ in the AFM phase. (e) With increased Mn vacancies, $H_{IL} < H_D$ in the FM phase.